\documentclass[aps,prb,showpacs,twocolumn]{revtex4}

\usepackage{graphicx}

\def\bra{\langle}
\def\ket{\rangle}
\def\vk{{\bf k}}

\begin{document}

\title{Impurity induced low-energy resonances in Bi$_2$Sr$_2$CaCu$_2$O$_{8+\delta}$}
\author{Jian-Ming Tang}
\author{Michael E. Flatt\'e}
\affiliation{Department of Physics and Astronomy, University of Iowa, Iowa City, Iowa 52242-1479}


\begin{abstract}
  
  We study sharp low-energy resonance peaks in the local density of
  states (LDOS) induced by Zn impurities or possible Cu vacancies in
  superconducting Bi$_2$Sr$_2$CaCu$_2$O$_{8+\delta}$.  The measured
  structure of these near-zero-bias resonances is quantitatively
  reproduced by an extended impurity potential without invoking
  internal impurity states or sophisticated tunneling models.  The Zn
  potential extends at least to the nearest-neighbor Cu sites, and the
  range of order parameter suppression extends at least $8$~{\AA} away
  from the Zn site.  We further show that the local spin
  susceptibilities near Zn impurities increase rather than decrease
  with decreasing temperature in the superconducting state due to the
  sharp increase of LDOS near the Fermi level.

\end{abstract}

\pacs{74.62.Dh,74.25.Jb,74.20.Rp,74.72.Hs}

\maketitle

Recent scanning tunneling microscopy (STM) measurements on the
high-$T_c$ superconductor Bi$_2$Sr$_2$CaCu$_2$O$_{8+\delta}$ (BSCCO)
have revealed the detailed spectral and spatial structures of sharp
resonances inside the superconducting gap induced by the impurities
Zn~\cite{Pan2000} and Ni~\cite{Hudson2001}, and by defects such as Cu
vacancies.\cite{Hudson2003} The properties of these resonances depend
on interactions at the atomic scale between impurity potentials and
the coherent superconducting state. As work progresses to probe these
defects in the underdoped regime, and above $T_c$, it is essential to
understand whether the properties in the superconducting state can be
understood within $d$-wave BCS theory, or whether vestiges remain of
the exotic correlated state present at higher
temperatures.\cite{Flatte2000,Polkovnikov2001,Zhu2001,Martin2002}
Detailed theoretical studies have shown that the resonances induced by
a Ni impurity in BSCCO can be well described by Bogoliubov
quasiparticles scattering off a potential in a $d$-wave superconductor
with a small, but discernible, local order parameter
suppression.\cite{Tang2002} On the other hand, theories based on
pointlike potentials have failed to explain quantitatively both the
spatial structure of the resonances and the equally important
information yielded by the spatial structure of the ``coherence
peaks'', either for Zn impurities or Cu
vacancies.\cite{Flatte2000,Polkovnikov2001,Zhu2001,Martin2002}

Zn impurities induce a sharp resonance at an energy close to the Fermi
level. Experimental results show that these resonances possess a large
spectral weight of local density of states (LDOS) at the impurity
site, whereas theories based on pointlike potentials predict that the
spectral weight at the impurity site is strongly suppressed by the
unrealistically large strength of the potential. This discrepancy in
the spatial structure has led to various proposals.  Some suggest the
need for strongly correlated models, in which the internal states of
the impurity are important,\cite{Polkovnikov2001,Zhu2001} and others
suggest that the spatial structure measured by STM is filtered by the
surface Bi-O layer above the Cu-O planes.\cite{Zhu2001,Martin2002}
However, none of these proposals has provided a coherent quantitative
account for the STM spectra, at both the resonance and the coherence
peaks, for both Zn and Ni impurities. In this Letter, we show that the
effects of Zn impurities over the entire spectrum can be
quantitatively described by spatially extended potentials, and,
therefore, render the potential model a fully quantitative description
for both Zn and Ni impurities in BSCCO. A principal, and unexpected,
conclusion is that the potentials required to model the Zn impurity
are nonmagnetic and relatively weak ($< 100$~meV), comparable in
strength to the potentials induced by the Ni impurity. Another
important result is that the order parameter suppression near Zn is
much more extensive than for Ni. We further report that the Cu
vacancy~\cite{Hudson2003} is well described within such a potential
model, thus demonstrating that the potential model works for all three
known defects probed by STM in BSCCO.

The appearance of near-zero-bias resonances is intrinsically important
because they greatly influence the bulk response to external fields,
such as conductivity and spin susceptibility.  Here we also calculate
the local spin susceptibilities near a Zn impurity.  In the
homogeneous bulk superconducting state, the spin susceptibility
decreases with decreasing temperature due to the opening of a gap in
LDOS around the Fermi level. However, in the presence of a sharp
near-zero-bias resonance, the local spin susceptibilities near the
impurity increase with decreasing temperature, and eventually the
values peak at a low temperature scale set by the resonance energy. In
other words, above this temperature scale, the local spin
susceptibilities inversely scale with temperature, similar to the
Curie behavior of a paramagnetic ion in an insulator. We find such
behavior describes the Knight shifts on and near Zn in BSCCO.

\begin{figure}

\centerline{ \includegraphics[width=\columnwidth]{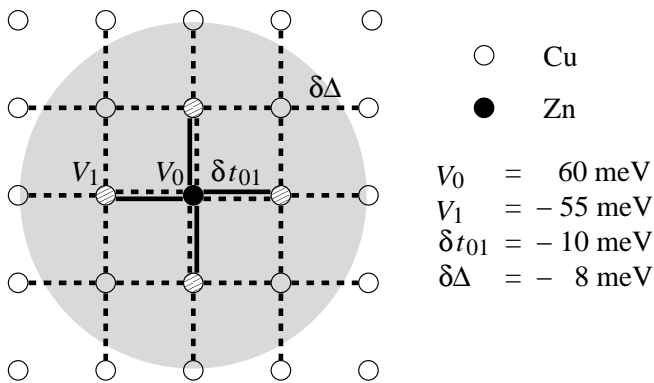} }

\caption{ A schematic diagram showing the proposed model of a Zn
  impurity on the Cu--O plane. Parameters extracted from fitting to
  the data are listed on the right. The superconducting order
  parameters on the dashed links are completely suppressed. This model
  also works for the Cu vacancy with $V_0 = 110$~meV, $V_1 = -50$~meV,
  and $\delta t_{01} = 19$~meV. For the Cu vacancy the order parameter
  was suppressed to zero ($\delta\Delta_{ij} = -\Delta_{ij}$) within
  8~{\AA} of the vacancy, and by 70\% ($\delta\Delta_{ij} =
  -0.7\Delta_{ij}$) from 8~{\AA} to 12~{\AA} from the vacancy. }

\label{fig:model}

\end{figure}

\begin{figure}

\centerline{ \includegraphics[width=\columnwidth]{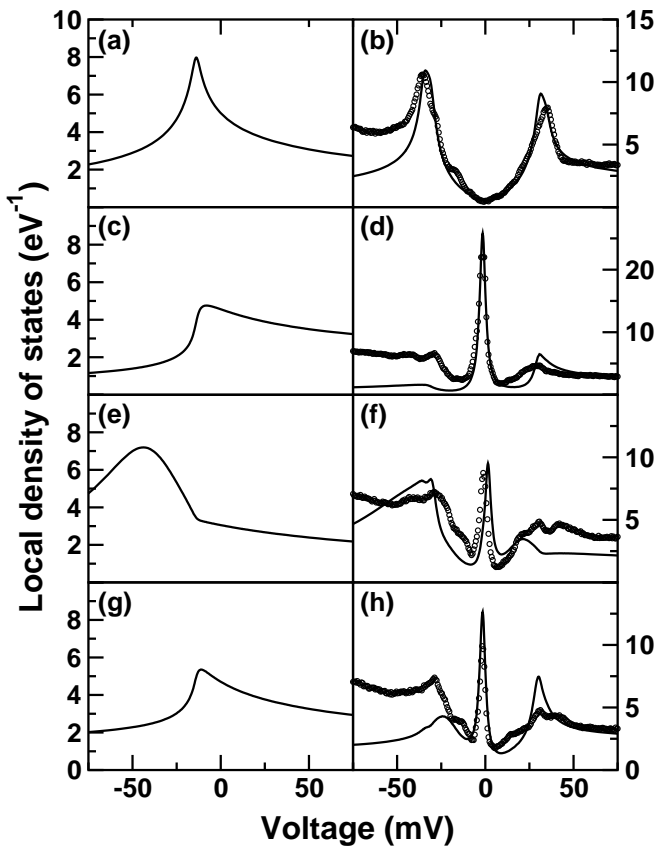} }

\caption{ LDOS spectra per unit cell at various sites near a
  Zn impurity. The left panels show the spectra in the ``normal
  state'' (simply setting $\Delta_{ij}=0$) using the same impurity
  parameters. The right panels show the spectra in the superconducting
  state. Solid lines show the calculated results. (a) and (b) are the
  spectra at a site far away from the Zn impurity. (c) and (d) are the
  spectra right at the impurity site. (e) and (f) are the spectra at
  the 1st nearest-neighbor sites. (g) and (h) are the spectra at the
  2nd nearest-neighbor sites. Open circles ({\large $\circ$}) show the
  STM differential conductance data.\cite{Pan2000} The data was
  rescaled by a constant factor identical for all the spectra. Note
  that the effective potential in the ``normal state'' could be very
  different.  }
\label{fig:spectra}
\end{figure}

\begin{figure}

\centerline{ \includegraphics[width=\columnwidth]{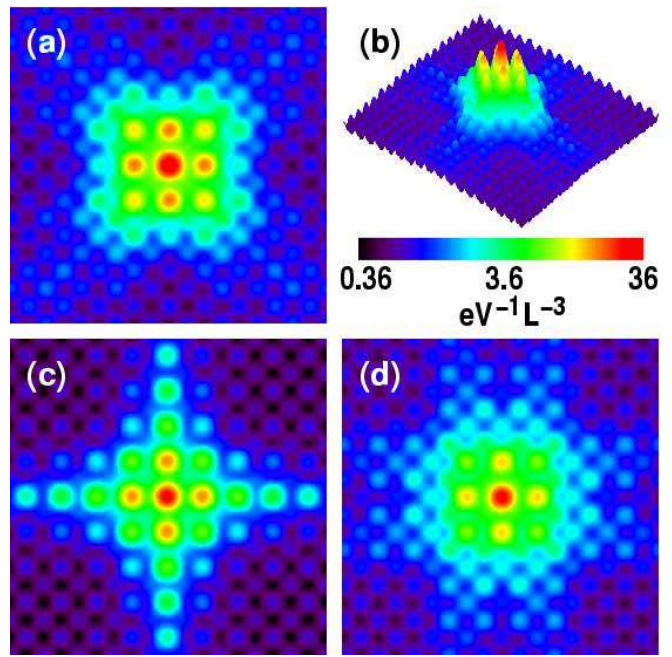} }

\caption{ (color)
  Spatial structure of the LDOS near a Zn impurity at $-2$ meV in (a)
  2D and (b) 3D view. We have spatially distributed the LDOS at each
  unit cell according to a normalized Gaussian with a width of half
  the lattice spacing ($L$).  The LDOS is shown in logarithmic scale.
  The Cu-O lattice is rotated by $45^\circ$ so that the horizontal
  axis is aligned with the superlattice modulation. To demonstrate the
  effect of the band structure, we carried out similar calculations
  for tight-binding bands with (c) only the nearest-neighbor hopping
  ($\{t_1,\mu,V_0,V_1\}=\{150, 10, 100, -100\}$ meV) and with (d)
  only the nearest-neighbor and the second-nearest-neighbor hopping
  ($\{t_1,t_2,\mu,V_0,V_1\}=\{150, -40, -150, 75, -75\}$ meV).
  The chemical potential is chosen so that the Van Hove singularity is
  fixed at the same energy. The impurity potential is also slightly
  altered to keep the resonance energy at the same place. }

\label{fig:spatial}

\end{figure}

We begin by describing the nature of the spatially-extended potential
that quantitatively describes all of the local spectra near the Zn
impurity (shown in Fig.~\ref{fig:model}). The failure of a pointlike
potential to induce the correct spatial structure of Zn's
near-zero-bias resonance originates from the large value of the
potential ($>1$~eV) required to draw states deep into the center of
the gap.  The strong on-site potential diminishes the spectral weight
of the LDOS at the impurity site, and moves the largest spectral
weights to the four nearest-neighbor sites.  Experimentally the
largest spectral weights, in order from largest to smallest, are at
the impurity site, at the second-nearest-neighbor sites, and then at
the third-nearest-neighbor sites (See Figs.~\ref{fig:spectra} and
\ref{fig:spatial}). Such a structure is consistent with a non-zero
potential at the four nearest-neighbor sites.  A key feature is that
the potential at the impurity site is of the opposite sign to those at
the nearest-neighbor sites, consistent with the structure of a
screened charge perturbation. The presence of a nonmagnetic spatially
extended potential and a short-ranged charge oscillation is consistent
also with the results of nuclear quadruple resonance (NQR)
measurements on Zn impurities in YBa$_2$Cu$_4$O$_8$
(YBCO).\cite{Williams2001} These potentials at the nearest-neighbor
sites are much more effective than merely an on-site potential in
inducing a sharp resonance close to the Fermi level. Unrealistically
large values for the potential are no longer required. A
moderate-range local suppression of the order parameter also further
drives the resonance energy toward the Fermi
level.\cite{Shnirman1999,Tang2002} Thus a quantitative understanding
of the experimental data requires careful consideration of the
detailed potential structure. The same model, but with different
values for the potentials, including a longer-range order parameter
suppression, describes the Cu vacancy.

We use the Koster-Slater approach to calculate the LDOS around
impurities.\cite{Flatte1997b,Flatte1999,Tang2002} Homogeneous BSCCO is
modeled by a one-band tight-binding Hamiltonian, as described in
Ref.~\onlinecite{Tang2002}.  The Green's function of the homogeneous
system is first evaluated with an energy resolution $\delta$. The
Green's function with impurities is then calculated by solving the
Gorkov (Dyson) equation. The impurity potential consists of
site-diagonal matrix elements at the impurity site $(V_0)$ and at the
nearest-neighbor sites $(V_1)$, and off-diagonal matrix elements
corresponding to modifications of the hopping matrix elements ($\delta
t$), and of the superconducting order parameters ($\delta\Delta$) as
described in Ref.~\onlinecite{Tang2002} for the Ni impurity. The full
Hamiltonian that describes the one impurity problem takes the
following form,
\begin{eqnarray}
H & = & -\sum_{\bra i,j\ket,\,\sigma}(t_{ij}+\delta t_{ij})c^\dagger_{i\sigma}c_{j\sigma} \nonumber\\
&& +\sum_{\bra i,j\ket}\left[(\Delta_{ij}+\delta\Delta_{ij})c^\dagger_{i\uparrow}c^\dagger_{j\downarrow}+{\rm H.c.}\right] \nonumber\\
&& +\sum_{\sigma}\left[V_{0}c^\dagger_{0\sigma}c_{0\sigma}+V_1\sum_{k=1}^4c^\dagger_{k\sigma}c_{k\sigma}\right] \;,\label{eq:H}
\end{eqnarray}
where $i$ and $j$ label the lattice sites (the impurity resides at
site 0), and $\sigma$ labels spin. The hopping matrix elements,
$t_{ij} = \{148.8, -40.9, 13, 14, -12.8\}$ meV, are based on a
one-band parameterization of the angle-resolved photoemission
data.\cite{Norman1995} The superconducting order parameters,
$\Delta_{ij}$, of the homogeneous system are only non-zero on the
bonds connecting two nearest-neighbor sites, $\Delta_{i,i+\hat
  x}=-\Delta_{i,i+\hat y}=\Delta_0/4$, where $\Delta_0$ is the gap
maximum. The momentum-dependent order parameter resulting from these
$\Delta_{ij}$ has $d$-wave symmetry, $\Delta_\vk=(\Delta_0/2)(\cos
k_xL-\cos k_yL)$, where $L$ is the lattice spacing between two Cu
atoms.

To construct the effective potential for the Zn impurity, we relate
important features in the LDOS data to different parts of the
potential structure.  First of all, it is known from STM that there is
microscopic inhomogeneity of the LDOS gap, which appears to be related
to the local doping concentration.\cite{Pan2001} Judging by the
experimental LDOS spectrum at a site within the same local patch as
the Zn impurity, but reasonably far away from the impurity, we
determine that the gap maximum $\Delta_0$ is about $32$ meV in
Ref.~\onlinecite{Pan2000}. The local chemical potential $\mu$ is
shifted with respect to the optimal doping value
($-130.5$ meV)\cite{Norman1995} by $-20$ meV, which
sets the Van Hove peak to be about $14$ meV below the Fermi level. The
energy resolution $\delta$ is about $2$ meV, roughly judging by the
widths of the coherence peaks and of the Zn resonance. The range of
order parameter suppression is then determined by the closing in of
the local gap edge.  Because the amplitude of the remnant coherence
peak is strongly suppressed and the appearance of mid-gap resonances
contributes additional spectral weight inside the gap, our fit for the
gap edge is not as quantitative as for other features of the
resonance. Nevertheless, the minimum range of the order parameter
suppression is determined to be about 2 lattice constants away from
the impurity.  That is, the order parameter is completely suppressed
within a circle with a radius of approximately $8$~{\AA}.  This is a
much more extensive order parameter suppression than seen around the
Ni impurity,~\cite{Tang2002} and suggests that Zn is more destructive
to local superconductivity.

The matrix elements $V_0$, $V_1$ and $\delta t_{01}$ are then
determined based on the resonance energy and the peak amplitudes at
three different sites (the impurity site, the 1st and the 2nd
nearest-neighbor sites). Although all 3 parameters jointly determine
the spatial structure, $V_1$ is the most dominant component.  $V_0$ is
used to obtain the correct resonance energy and $\delta t_{01}$ is
used to adjust the ratio between the peaks.  Our procedure provides
similar spatial structures if the sign of both $V_0$ and $V_1$ are
reversed -- we can only determine that they have opposite sign
relative to each other.\cite{secondset}  Motivated by the NQR
experiments that suggest electron charge accumulation on the
nearest-neighbor sites of Zn in YBCO, we choose the potential energy
on the nearest-neighbor sites to have negative values (attractive to
electrons).

Figure \ref{fig:spectra} shows the detailed comparisons of our
calculations with data. In general, we have a good fit for the spectra
in a wide energy range inside the gap. The fit is less satisfactory
for the energy range beyond the coherence peaks. Therefore, the
determination of the gap edge is not as precise as the determination
of other parameters based on the resonance energy and peak spectral
weights. We note that for every midgap resonance one expects a feature
on the nearest-neighbor sites which is on the opposite side from that
on the impurity site itself.\cite{Flatte1998,Flatte2000} For these
spatially-extended potentials, however, the spectral weight of these
resonance peaks is highly reduced and likely not visible.

The spatial structure of the resonance peak, suitable for direct
comparison with Ref.~\onlinecite{Pan2000}, is shown in
Fig.~\ref{fig:spatial}.  Note that the spectral weight is most peaked
along the $(1,1)$ direction of the Cu-O lattice in the near region
around the impurity.  Far from the impurity, the tails align with the
Cu-O bond direction (This is also the case for Ni, but the signal is
much weaker.) As our fit of the local potentials was constrained
entirely by features near the impurity, obtaining good agreement with
measured data for this orientation of the resonance tail suggests that
we have a reasonable description of the BSCCO band structure near the
Fermi level. For example, in a simple nearest-neighbor tight-binding
model we find that the long-ranged tail of the resonance incorrectly
follows the diagonal of the Cu-O lattice [Fig.~\ref{fig:spatial}(c)].
The tail begins to rotate to follow correctly the Cu-O bond direction
when the second-nearest-neighbor hopping is added
[Fig.~\ref{fig:spatial}(d)], and the proper orientation is obtained in
the full tight-binding model for BSCCO. We have verified that this
defect model, with the different potentials reported in
Fig.~\ref{fig:model}, also describes well the more limited Cu vacancy
data in Ref.~\onlinecite{Hudson2003}. The order parameter suppression
even more extended around the Cu vacancy than around the Zn impurity;
our model parameters are listed in the Fig.~\ref{fig:model} caption.

\begin{figure}

\centerline{ \includegraphics[width=\columnwidth]{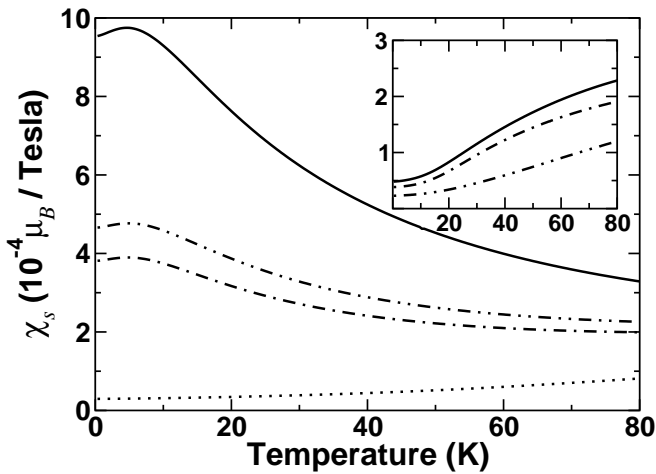} }

\caption{ The local spin susceptibilities at the Zn site (solid line),
  at the 1st nearest-neighbor sites (dot-dashed line), at the 2nd
  nearest-neighbor sites (double dot-dashed line), and at a faraway
  site (dotted line). The susceptibilities near Zn show a Curie-like
  behavior in an intermediate temperature range (above $10$ K and
  below the superconducting transition temperature).  The inset shows
  the corresponding susceptibilities near a Ni impurity, which simply
  increase with temperature. }
\label{fig:susceptibility}
\end{figure}

To show that the bulk response to fields can be dramatically changed
by the appearance of the near-zero-bias resonance, we calculate the
local spin susceptibilities (suitable for determining the Knight
shifts) around the impurity using our results for the LDOS.  The local
spin susceptibility at site $j$ is defined as
\begin{eqnarray}
\chi_s & = & \frac{\mu_B^2}{k_BT} \int d\omega A_j(\omega)f(\omega)[1-f(\omega)] \;,
\end{eqnarray}
where $\mu_B$ is the Bohr magneton, $k_B$ is the Boltzmann constant,
$T$ is temperature, $A_j(\omega)$ is the LDOS at site $j$, and
$f(\omega)$ is the Fermi-Dirac distribution function. We assume $g=2$.
The results for both Zn and Ni impurities are shown in
Fig.~\ref{fig:susceptibility}.  For Zn the local spin susceptibility
increases as the temperature decreases, and there is a maximum at a
very low temperature, which is determined by the resonance energy.
The strongest response is at the impurity site,
and the signal strengths at the 1st and 2nd nearest-neighbor sites are
approximately equal. Ni, which has a much higher resonance energy,
does not show this low-temperature maximum, and thus does not appear
to be paramagnetic in this type of experiment. Thus we find that the
resonance induced by a completely nonmagnetic impurity potential for
Zn generates a signal in spin susceptibility experiments that closely
mimics the expected signal from a free spin (as suggested in
Ref.~\onlinecite{Tallon2002}).  This result may have implications for
measurements of spin susceptibilities in
YBa$_2$Ca$_3$O$_{7-\delta}$.\cite{Bobroff2001}

We have provided a reasonable nonmagnetic potential model that
quantitatively reproduces the spatial structure of the Zn impurity
resonance and the spatial structure of the coherence peaks near Zn. We
have verified that the potential model also produces accurate results
for the Cu vacancy. The model is based on the same BSCCO electronic
structure as used previously to describe Ni, thus demonstrating that a
consistent picture is possible that quantitatively describes all three
defects in BSCCO.  Surprisingly small effective potentials are
required to describe the Zn resonance, even though it occurs near zero
energy. The order parameter suppression we find near Zn is more
extensive than near Ni, but not as extensive as near the Cu vacancy.
We have also demonstrated that the resonance near Zn would respond in
spin susceptibility measurements in a way that would mimic a free
spin, whereas Ni would not. Finally, the potential for the vacancy,
and to a lesser extent Zn, scatters quasiparticles most effectively
for momentum transfer $(\pi,\pi)$. This may broaden quasiparticle
signatures significantly at the $(\pi,0)$ points of the Brillouin
zone. These results provide a foundation for better understanding of
measurements that are likely to be made on these defects in the more
exotic ``pseudogap'' state of BSCCO at higher temperatures.

We thank E.~W. Hudson and J.~C. Davis for providing data shown in
Fig.~\ref{fig:spectra}.  This work is supported by ONR Grant
Nos.~N00014-04-1-0046 and N00014-99-1-0313.


\begin{thebibliography}{15}
\bibitem{secondset} The STM data for Zn impurities can also be fitted using
$V_0=-85$ meV and $V_1=60$ meV, and the other parameters are the same.
\end{thebibliography}
\end{document}